\documentclass[prl,aps,twocolumn]{revtex4}
\usepackage{graphicx}
\usepackage{color}
\begin{document}

\renewcommand{\thefootnote}{\fnsymbol{footnote}}
\newcommand{\redc}[1]{{\color{red}#1}}
\newcommand{\bluec}[1]{{\color{blue}#1}}
\draft
\title{Time-Resolved Cavity Nano-Optomechanics in the 20-100 GHz range}

\author{S. Anguiano$^{1}$, A. E. Bruchhausen$^{1}$, B. Jusserand$^2$, I. Favero$^3$, F. R. Lamperti$^{3,4}$, L. Lanco$^4$, I. Sagnes$^4$, A. Lema\^itre$^4$, N. D. Lanzillotti-Kimura$^4$,  P. Senellart$^4$,
and A. Fainstein$^{1, }$\footnote{email:afains@cab.cnea.gov.ar}}
\affiliation{$^1$Centro At\'omico Bariloche \& Instituto Balseiro, C.N.E.A., CONICET, 8400 S. C. de Bariloche, R. N., Argentina}
\affiliation{$^2$Institut des NanoSciences de Paris, UMR 7588 C.N.R.S. - Universit\'e Pierre et Marie Curie, 75015 Paris, France}
\affiliation{$^3$Universit\'e Paris Diderot, Sorbonne Paris Cit\'e, Laboratoire Mat\'eriaux et Ph\'enom\`enes
Quantiques, CNRS-UMR 7162, 10 rue Alice Domon et L\'eonie Duquet, 75013 Paris, France}
\affiliation{$^4$Centre de Nanosciences et de Nanotechnologies, C.N.R.S., Univ. Paris-Sud, Universit\'e Paris-Saclay, C2N Marcoussis, 91460 Marcoussis, France}



\maketitle

{\bf Applications of cavity optomechanics span from gravitational wave detection~\cite{Ligo} to the study of quantum motion states in mesoscopic
mechanical systems.~\cite{ReviewCOM} The engineering of resonators supporting strongly interacting mechanical and optical modes is central to these developments. However, current technological and experimental approaches limit the accessible mechanical frequencies to a few GHz,~\cite{Favero,Painter,Tang,Baets} imposing hard constraints on quantum mechanical studies.~\cite{O'Connell,Teufel,Chan,Regal} Here we demonstrate the optical control of 20-100~GHz mechanical modes confined in the three dimensions within semiconductor nano-optomechanical pillar cavities. We use a time-resolved transient optical reflectivity technique and access both the energy spectrum and dynamics of the mechanical modes at the picosecond timescale. A strong increase of the optomechanical coupling upon reducing the pillar size is observed together with unprecedent room temperature Q-frequency products above $10^{14}$. The measurements also reveal sideband generation in the optomechanical response. Such resonators can naturally integrate quantum wells and quantum dots,~\cite{LPNQDs} enabling novel applications in cavity quantum electrodynamics and high frequency nanomechanics.~\cite{Restrepo,Kyriienko,Rozas_Polariton}}

Solid state cavity optomechanical devices include microspheres, toroidal microcavities, membranes, MEMS, and planar photonic waveguides, among others.~\cite{ReviewCOM} Light-induced rigidity, optomechanical cooling down to the quantum ground state of mechanical motion,~\cite{O'Connell,Teufel,Chan,Regal}  and optomechanical self-oscillation~\cite{Karrai,Kippenberg,Zhao,Grudinin} are demonstrated consequences of the dynamical back-action in these systems. Typical operation frequencies range from a Hertz to a few GHz, the latter records having been demonstrated in microdisk resonators and optomechanical crystals~\cite{Favero,Painter,Tang,Baets} limited essentially by nano-fabrication techniques such as lithography as well as the lack of suitable detection methods.

A promising way to confine mechanical motion is to use layers of materials presenting different acoustic impedance so as to define acoustic Bragg mirrors. This can be done using semiconductor materials like GaAs and AlAs that present a strong acoustic impedance contrast \cite{FainsteinPRL2013} and  can be grown by molecular beam epitaxy with sub-atomic size accuracy, one or two orders of magnitude better than the best nanofabrication techniques. In the present work, we study GaAs/AlAs nano-optomechanical resonators in the form of  pillar cavities based on Bragg structures that perform both as acoustic and optical reflectors confining both light and the mechanical vibrations in the three spatial dimensions. We demonstrate the optomechanical coupling of mechanical modes in the 20-100 GHz frequency range with near infrared light in these nanostructured pillars. The mechanical modes are evidenced through pump-probe optical techniques that allow measuring the dynamical optomechanical response at the picosecond timescale. Record room temperature Q-frequency products above $10^{14}$ are measured as well as a strong increase of the optomechanical coupling strength  when reducing the pillar lateral size.

A  GaAs/AlAs planar acoustic cavity of wavelength  $\lambda$ can be formed by growing a $\lambda/2$-thick  spacer between two distributed Bragg reflectors (DBRs). Each DBR is composed by periodically alternating $\lambda/4$ layers. The GaAs/AlAs acoustic impedance contrast and the number of layers determine the acoustic quality factor of the mode.
It was recently shown that the contrast in the acoustic impedances of GaAs and AlAs $(0.83)$ equals the contrast in the optical indices of refraction $(0.83)$. As a result,  an optimized planar acoustic microcavity is automatically an optical cavity confining photons of the same wavelength,  determined by the spacer thickness.~\cite{FainsteinPRL2013} Because of the strong  difference between the light and sound speeds,  a microcavity confining near infrared photons confines mechanical vibrations in the 20-100 GHz range. Such planar cavity can be etched  to obtain pillars with micron-size diameters. In the optical domain, such structures are known to confine light in the three directions. We show now that such  pillar structures also perform as nano-optomechanical resonators with unprecedented high frequencies.

\begin{figure*}
    \begin{center}
    \includegraphics[scale=0.45,angle=0]{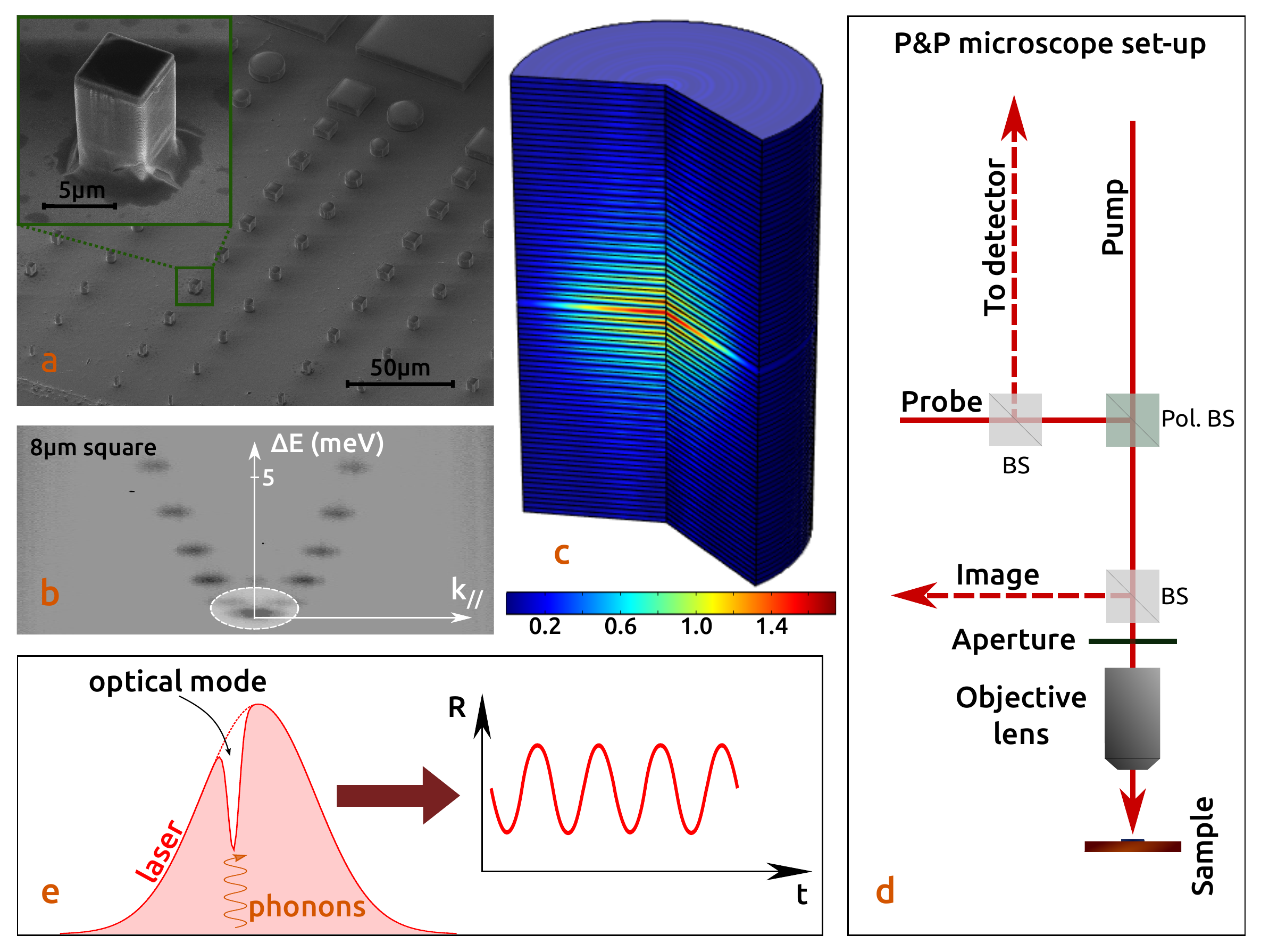}
    \end{center}
    \vspace{-0.4 cm}
\caption{{\bf Pillar nano-optomechanical resonators and experimental technique} (a) SEM images of the studied micron-sized pillar structures. An array of circular and square pillars with lateral sizes ranging from 50 to 1 micron is shown. The inset presents a zoom on a 5 micron square pillar. (b) k-space image of the optical cavity modes for a square pillars of 8 micrometer lateral size. The shaded ellipse represents the profile (energy broadening and angular dispersion) of the pump and probe laser pulses. (c) Spatial distribution of the absolute value of the volumetric strain ($dV/V$) associated with a confined mechanical mode around 20 GHz, calculated using finite element methods. (d) Scheme of the ultrafast resonant laser micro-spectroscopy set-up. (e) Scheme of the process leading to mechanical signals in reflectance difference time resolved spectroscopy.
\label{fig01}}
\end{figure*}

We study opto-mechanical pillar cavities consisting of a $\lambda/2$ GaAs spacer cavity sandwiched between Ga$_{0.9}$Al$_{0.1}$As/Ga$_{0.05}$Al$_{0.95}$As $\lambda/4$ DBRs. Figure~\ref{fig01}(a) presents a scanning electron microscopy image of an array of circular and square pillars with lateral sizes ranging from 50 to $1~\mu$m.  The inset presents a zoom on a $5~\mu$m square pillar, evidencing the precision of the fabrication process. Results presented in this paper correspond to square shape pillars of varying lateral size $L$. The opto-mechanical resonators have been designed to confine photons around 870 nm and mechanical modes in the $\Omega \approx 20$~GHz range, with higher order modes at frequencies given by $(2p+1) \Omega$ ($p$ being an integer number). Even order gaps are closed in $\lambda/4,\lambda/4$ mirrors, preventing the existence of the corresponding cavity modes.

The full 3D confinement of the optical modes~\cite{JMGerard} can be evidenced by a photoluminescence k-space mapping of the modes, as illustrated for a $8 \mu$m pillar in Fig.~\ref{fig01}(b). Lateral confinement leads to a discretization of the parabolic dispersion of the cavity mode, with mode spacing increasing when reducing the pillar lateral size.~\cite{JMGerard_pillarDispersion}

Full 3D mechanical confinement is also expected from the boundary conditions at the pillar edge where $\tensor{\sigma} \cdot \vec{n} = \vec{0}$, with $\tensor{\sigma}$ the stress tensor and $\vec{n}$ the normal to the surface. Finite elements methods indeed predict confined mechanical modes, as illustrated in Fig.~\ref{fig01}(c) showing the absolute value of the volumetric strain $|dV/V|$ corresponding to a confined vibration mode around 20~GHz.

Ultrasensitive motion measurement techniques, standard in cavity optomechanics, are not readily available to  evidence such high frequency mechanical modes. To do so we implemented a time-resolved differential optical reflectivity (pump-probe) measurement~\cite{Thomsen,Bartels} with  micrometer lateral spatial resolution. A scheme of the set-up is presented in Fig.~\ref{fig01}(d). This technique gives full access to the response of the mechanical resonator in a forced oscillator regime, both in the temporal (picosecond resolution) and spectral domains.  A first picosecond optical pulse interacts with the sample,  generating coherent mechanical vibrations that modulate both the shape and the indices of refraction of the layers forming the pillar; then a delayed probe pulse measures the instantaneous optical reflectivity of the sample.
This is schematically shown in Fig.~\ref{fig01}(e). The envelope curve is the spectrum of the incoming probe beam, whereas the red shaded curve describes the spectrum of the reflected probe beam. The time-dependent measured signal (represented as R in the figure) is the total reflected probe power, i.e. the red area. Its instantaneous value directly depends on the energy of the optical cavity mode (appearing as a dip in the figure). The laser pulses are $\sim 1.5$~meV wide, while the optical cavity mode prior to the optical excitation is $\sim 0.7$~meV wide. By changing the delay between the pump and probe pulses we reconstruct the time-evolution of the optical reflectivity modulated by the mechanical vibrations.  For the case of GaAs/AlAs-based pillars, one can profit from an additional resonant enhancement of the involved optomechanical interaction - (in addition to the standard radiation pressure) by working close to the GaAs electronic gap at $\approx 870$~nm.~\cite{Favero,FainsteinPRL2013}  At these wavelengths, the change in the dielectric function of GaAs induced by the confined phonons is orders of magnitude stronger than far from the gap.~\cite{JusserandPRL2015}

Two main effects are induced by the pump pulse. Through two-photon and free carrier absorption a very fast (around 1~ps) modification of the GaAs index of refraction blue-shifts the optical cavity mode by a few meV, spanning a large part of the laser pulse bandwidth. Equilibrium is recovered through carrier recombination typically within 1-4~ns.~\cite{JMGerard_pillardynamics} We will refer to the reflectivity variations induced by this transient as the {\em electronic} contribution.  In addition, the excited carriers generate coherent mechanical excitations through the  deformation potential mechanism (or ``optoelectronic force'').~\cite{RuelloUltrasonics}  Photoexcited carriers imply a sudden change of energy landscape for the ion cores, that subsequently start to move around the new equilibrium position. These coherent vibrations come with a displacement of the interfaces, and modulate the index of refraction of the different layers in the device through the photoelastic mechanism.~\cite{FainsteinPRL2013,Baker,Rakich} We  refer to the latter as the {\em mechanical} contribution (illustrated in Fig.~\ref{fig01}(e)). This all results in a modulation of the cavity mode energy, leading to a time dependent variation of the reflectivity that is monitored by the probe pulse.

\begin{figure}
    \begin{center}
    \includegraphics[scale=0.7,angle=0]{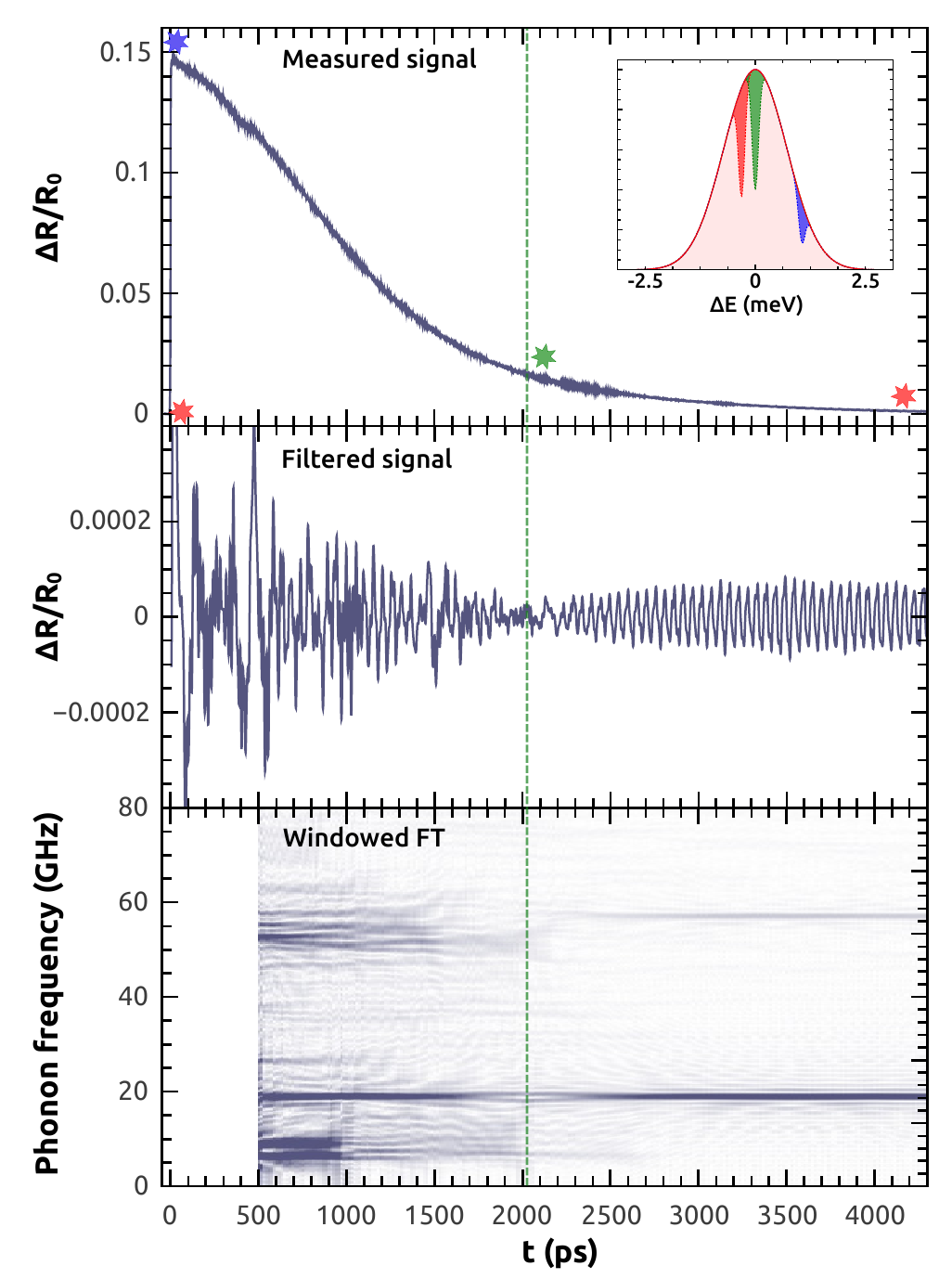}
    \end{center}
    \vspace{-0.8 cm}
\caption{{\bf  Time resolved transient  optomechanical measurement in the 5-100 GHz range}  Top: As-measured differential reflectance time trace obtained for a 5 micrometer square pillar, varying the delay $t$ between the pump and probe pulses. The inset is a schematic representation of the laser pulse energy distribution (shadowed gaussian profile), and the position of the cavity mode (appearing as a dip in the reflected probe) at different times after pump excitation. The time before the arrival of the pump is indicated with the initial red star and dip, after pump excitation with blue star and dip. Excited carrier relaxation leads to the faster recovery of the cavity mode, which around 2000~ps passes through the central energy of the laser (green star and dip). At longer times determined by carrier and thermal relaxation the equilibrium situation is recovered (final red star and dip). Middle: filtered time trace corresponding to frequencies between 5 and 100~GHz. The vertical dashed line indicates the delay corresponding to zero laser-cavity-mode detuning. Bottom panel: Windowed Fourier transform (WFT) of the filtered trace, obtained with 1000~ps windows.
\label{fig02}}
\end{figure}

Figure~\ref{fig02} presents a typical signal measured in a pump-probe experiment performed on a $5 \mu$m pillar. The top panel displays the as-measured reflectance difference trace varying the delay $t$ between the pump and probe pulses. This signal is mainly determined by the electronic contribution. Several snapshots describing the relative position of the laser spectrum and the probed optical cavity mode are shown for different delays in the top panel in Fig.~\ref{fig02}, using the same colors as the stars marking the associated time delays in the measured differential reflectance trace. Red identifies the electronic equilibrium situation, corresponding to times prior to the pump excitation, and at long times when equilibrium is recovered. Blue indicates the instant just after pump excitation, and green the crossing through zero laser-cavity-mode detuning during the electronic relaxation transient. The middle panel shows the mechanical signal obtained by filtering out the slowly varying spectral components (frequencies smaller than 5~GHz), and leaving only frequencies up to 100~GHz characteristic of the mechanical contributions. Information concerning the time dependent spectral content of the mechanical signals is obtained from the windowed Fourier transform calculated using 1000~ps windows, shown in the bottom panel as a color intensity map. The center of the windows spans continuously from  $t=500$ to 4500~ps. Below 2~ns, the electronic excitation leads to a complex intermixed behavior between mechanical vibrations and the optical mode shift that we will discuss further below. We focus first on the delays above 2~ns where the fast electronic-induced effects vanish and the modulation is dominated by the coherent mechanical vibrations, i.e., on the clean oscillations and pure spectra evidenced in the middle and bottom panels of Fig.~\ref{fig02}.

\begin{figure}
    \begin{center}
    \includegraphics[scale=0.45,angle=0]{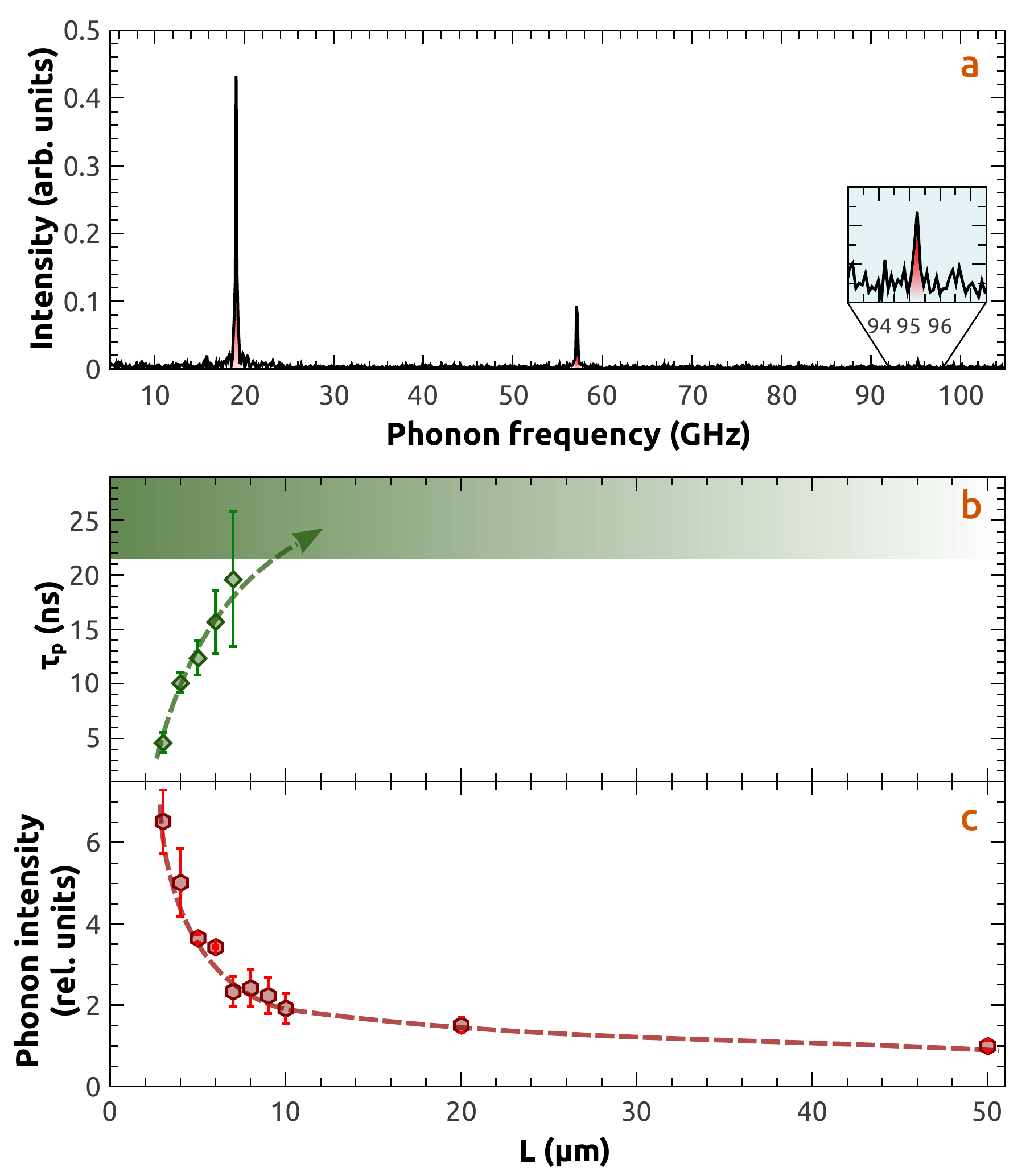}
    \end{center}
    \vspace{-0.4 cm}
\caption{{\bf  Confined optomechanical coupling in the 20-100 GHz range}  (a) Vibrational spectra corresponding to a 5 micron square pillar obtained through the Fourier transform of the oscillating part extracted from the reflectance difference signal (raw data shown in Fig.~\ref{fig02}).  The fundamental cavity mode at approximately 20~GHz and two overtones at 60 and 100~GHz are highlighted.
(b) Lifetime of the 20~GHz cavity confined mechanical vibration as a function of pillar size $L$, derived from the ratio of intensities measured at 4.5 and 12~ns. Error bars reflect the measured amplitude uncertainty. The colored region marks the resolution limit of our pump and probe technique. Dashed curves are guides to the eye.
(c) Mechanical signal intensity ($\Delta R/R_0$) corresponding to the 19~GHz confined acoustic mode, obtained from the Fourier transform of the time traces for varying square pillar size.
\label{fig03}}
\end{figure}

A typical spectrum for a $5 \mu$m square pillar obtained from the Fourier transform of the filtered trace between 2~ns and 10~ns is shown in Fig.~\ref{fig03}(a).  Three modes are clearly observed at $\sim 19$, $\sim 58$ and $\sim 95$~GHz, corresponding to the fundamental mode confined by the DBRs, and the third and fifth overtones, respectively. Because the pump induced perturbation is concentrated in the GaAs-spacer layer, the generated localized strain mainly corresponds to cavity confined vibrations. The symmetry of the optical mode results in a near-zero coupling to the mechanical modes with odd strain spatial distribution. Note that the technique allows to observe mechanical modes at very high frequency, in principle up to the THz range. The spectral resolution in our case is around 0.1~GHz, Fourier-transform limited by the length of the multiple-pass delay line used. This gives a lower bound for the measured room temperature mechanical Q-factors of the 19, 58 and 95~GHz modes of $\approx 200$, $\approx 600$, and $\approx 1000$, respectively, resulting in $Q \cdot f$ greater than $10^{14}$ which, to the best of our knowledge are record values at room temperature.  In practice, this implies that about a hundred coherent quantum control operations could be performed on this mechanical mode at room temperature.~\cite{ReviewCOM}

A blue shift of the mechanical modes could be expected when reducing the size of the micropillars due to lateral confinement effects, as in the optical case. For micro-size pillars the energy shift of the mode, compared to the planar structure, is expected to be smaller than 40 MHz,  below the experimental resolution. However, as shown now, the lateral confinement of the mechanical mode manifests itself both in the mechanical mode lifetime and the optomechanical signal intensity.

Figure~\ref{fig03}(b) shows the lifetime of the 19~GHz cavity confined acoustic vibration as a function of pillar size $L$, derived from the ratio of the intensity of the mechanical signal at 12~ns and 4.5~ns (see the Supplementary Information). Error bars reflect the measured amplitude uncertainty. The colored region marks the present resolution limit of our pump-probe technique, indicating that the lifetimes are at least of 20~ns. There is a clear  decrease of the mechanical mode lifetime for pillar sizes smaller than $7 \mu$m. In the optical domain, the Q-factor of pillar cavities generally decreases below a certain size depending on the etching and surface quality, because of sidewall losses.~\cite{GayralGerard} Dissipation induced by lateral surfaces is expected to become more relevant for smaller devices because of an increased surface to volume ratio, be it for photons or phonons. We note however that the measured lifetimes above 20 ns for pillar lateral sizes larger than $6 \mu$m are an indication of the high quality of the fabricated resonators. Indeed, this value is close to the phonon  lifetime in the GHz range studied in bulk GaAs/AlAs structures at room temperature. Note that  in contrast to photons, phonons are intrinsically anharmonic, meaning that they can decay into two or more phonons, or can scatter with background phonons. These three and four phonon processes are not significant at MHz frequencies, but become quite relevant in the GHz-THz range, and particularly at room temperature.~\cite{Rozas}

Figure~\ref{fig03}(c) shows the mechanical signal intensities corresponding to the fundamental 19~GHz confined mode, obtained from the Fourier transform of the time traces for varying square pillar size.
A strong increase of the mechanical signal with decreasing pillar size is observed. We interpret this behavior as evidence of an increase in the optomechanical coupling factor $g_0$  of the resonators, which determines the net effect a single phonon produces on the change in the optical reflectivity of the system.
The measured intensity of the mechanical signals depends on the pump intensity effectively exciting the pillar modes, in other words on how many photons actually interact with the GaAs spacer.  The phonon generation process saturates with the incident power.  We thus worked in a region of pump laser intensities that, due to this saturation, is not strongly dependent on the deposited power. The mechanical induced variation of the reflected laser power also depends on the amount of probe light impinging on and reflected from the pillar surface. To take this effect into account we normalized the time-dependent mechanical signal, $\Delta R$, with respect to the measured reflected continuous wave probe power, $R_0$ (see details of the procedure in the supplementary information).

Having demonstrated the three dimensional mechanical confinement in pillar cavities through the lateral size dependence of the intensity and lifetime of the observed confined modes, we discuss now the complex dynamics observed at shorter times in Fig.~\ref{fig02}. Pulsed studies in cavity optomechanics have only recently started to be reported as a means to access the transient backaction dynamics.~\cite{Painter_Transient} Here, our experimental method allows us to extract the mechanical transient dynamics of the nano optomechanical pillar cavities with picosecond resolution. A detailed description of the observed behavior is not yet available, but we can already draw some conclusions on the optomechanical behavior in this early stage where non-linearity plays an important role.

Important insight on the observed transient dynamics can be grasped from filtered traces and the windowed Fourier transform displayed in the middle and bottom panels of Fig.~\ref{fig02}. Strikingly, complex multiple frequency spectra are observed immediately after pump excitation and when the cavity mode rapidly evolves along the high-energy flank of the laser spectrum.~\cite{Splitting} The additional observed frequencies seem to concentrate more around the frequency of the two cavity confined modes at 19 and 58~GHz. The same systematics is observed in all the studied pillars, with lateral sizes ranging from 3 to $50~\mu$m. The generation of these complex spectral patterns strongly hints towards nonlinear interactions being at play in our resonators. In fact, the standard radiation pressure optomechanical coupling itself is nonlinear, and has been shown to lead to frequency doubling, multistabilities~\cite{Marquardt2006,Metzger} and chaotic behavior~\cite{TalCarmon,Vahala2007} under continuous-wave optical pumping. In contrast, our present experiments are carried in an ultra-fast time-resolved regime, directly revealing the transient at short times. Additionally, the laser detuning varies on a time-scale itself commensurable with mechanical periods, a situation that also greatly differs from conventional cavity optomechanics. These observations suggest that complex dynamics can emerge in these new situations. The generation of multiple frequencies that we witness at short time  resembles a  form of non-linear optomechanical back-action, but the fine modeling of these effects requires going beyond existing models of optomechanics.

The access to a new frequency range in three dimensionally confined optomechanical resonators removes some of the roadblocks in the quest of quantum phenomena and control at the single phonon level at high temperature.  The studied structures are naturally integrable with semiconductor quantum wells and quantum dots, opening the path to realizable polariton optomechanical resonators that synergically put into play the phenomena and application of cavity quantum electrodynamics with cavity optomechanics. Strongly enhanced interaction have been predicted in such devices, allowing cooling and control of quantum mechanical states at the single photon level. The manipulation and control of interactions of ultra high frequency mechanical modes in the 100~GHz range with polaritons and quantum emitters~\cite{Restrepo,Kyriienko,Rozas_Polariton} are only examples of new exciting prospects.

\section{Methods}

{\bf Sample.} Pillar microcavities are fabricated by etching a planar cavity structure into the shape of vertical posts.  The planar microcavity consists of a $\lambda/2$ GaAs layer enclosed by two distributed Bragg reflectors consisting of alternating Ga$_{0.9}$Al$_{0.1}$As/Ga$_{0.05}$Al$_{0.95}$As $\lambda/4$ layers, 28 pairs on the bottom, and 24 on top.  This epitaxialy grown planar structure had a gradient in thicknesses. The optical cavity mode could be thus tuned ($\pm 50$ meV) around the room-temperature GaAs 1s exciton transition $\approx 1.42$ eV, by changing the position of the laser spot on the sample. The optical Q-factor of the original planar structure is around $1.4 \times 10^4$.

{\bf Experiment.} Noise techniques usually used in cavity optomechanics are optimized to work in the telecommunication bandwidth range, taking advantage of all the already available developments for high frequency signal detection and treatment.  For the case of GaAs/AlAs-based pillars reported here, current state-of-the-art electronics working at ~900 nm wavelengths are typically limited to 5-10~GHz, and are consequently not applicable in the 20-100~GHz range.  The pump-probe technique has its own challenges because both pump and probe need to be degenerate and resonant with the pillar cavity mode. To avoid ``cross talking" of pump light into the detection channel both polarization and spatial filtering are required.  Pump and probe relative powers are controlled by a $\lambda/2$ retarder and a polarizing beam splitter. Laser pulses of approximately 1~ps duration are thus separated into crossed polarized pump and probe beams. The pump beam is modulated at MHz frequencies to allow for a synchonous detection using a lock-in amplifier, while the timing between pump and probe is set by passing the probe beam through a multi-pass 60~cm delay line. Pump and probe pulses are sent to impinge on the pillar surface through the same microscope objective. An aperture is introduced before the latter as spatial filter to limit the back-reflected pump light collected by the same microscope and sent to the detector.
Polarization filtering is attained by using first a polarizing beam splitter in the collection path, and then a second polarizer at the entrance to the detector.
The experiments were performed at room temperature with the laser wavelength set at 882~nm, close to and about 10~meV below the direct gap of the bulk GaAs constituting the cavity spacer. Due to this resonant coupling the GaAs cavity spacer acts as a spatially localized optomechanical transducer.  The optical cavity mode (spectrally narrower than the 1~ps laser pulses) is tuned slightly red-shifted respect to the center of the laser pulses (as shown in Fig.~\ref{fig01}(d)).
The resolution limit for the obtention of lifetimes is determined by the pulsed laser repetition rate (80~MHz, corresponding to 12.5~ns between pulses), and the use of the six-pass 60~cm mechanical delay line.

{\bf Acknowledgments:} This work was partially supported by the ANPCyT Grants PICT 2012-1661 and 2013-2047, the ERC Starting Grant No. 277885 QD-CQED, the French Agence Nationale pour la Recherche (grant ANR QDOM) the French RENATECH network, the Labex NanoSaclay,  and the international franco-argentinean laboratory LIFAN (CNRS-CONICET). N.D.L.K. was supported by the FP7 Marie Curie Fellowship OMSiQuD.

{\bf Author contributions:} The sample was fabricated by A.L.,  I.S. and P.S. The measurements and data processing were performed by S. A..  All authors participated to scientific discussions about the experimental data and manuscript preparation. The project was conducted by A.F.

{\bf Author information:} Correspondence and requests for materials should be addressed to afains@cab.cnea.gov.ar.


\begin{references}

\bibitem{Ligo} B. P. Abbott et al., GW150914: The Advanced LIGO Detectors in the Era of First Discoveries - LIGO Scientific and Virgo Collaborations, Phys. Rev. Lett. {\bf 116}, 131103 (2016).

\bibitem{ReviewCOM} M. Aspelmeyer, T. J. Kippenberg, and F. Marquardt, Cavity optomechanics, Rev. Mod. Phys. {\bf 86}, 1391 (2014).

\bibitem{Favero} L. Ding, C. Baker, P. Senellart, A. Lemaitre, S. Ducci, G. Leo, and I Favero, High Frequency GaAs Nano-Optomechanical Disk Resonator, Phys. Rev. Lett. {\bf 105}, 263903 (2010).

\bibitem{Painter} M. Eichenfield, J. Chan, R. M. Camacho, K. J. Vahala, and O Painter, Optomechanical crystals, Nature {\bf 462}, 78 (2009).

\bibitem{Tang} Xiankai Sun, Xufeng Zhang, and Hong X. Tang, High-Q silicon optomechanical microdisk resonators at gigahertz frequencies, Appl. Phys. Lett. {\bf 100}, 173116 (2012).

\bibitem{Baets} Rapha\"el Van Laer, Bart Kuyken, Dries Van Thourhout, and Roel Baets, Interaction between light and highly confined
hypersound in a silicon photonic nanowire, Nature Photonincs {\bf 9}, 199 (2015).



\bibitem{O'Connell} A. D. O'Connell, M. Hofheinz, M. Ansmann, R. C. Bialczak, M. Lenander, E. Lucero, M. Neeley, D. Sank, H. Wang, M. Weides, J. Wenner,
J. M. Martinis, and A. N. Cleland, Quantum ground state and single-phonon control of a mechanical resonator, Nature {\bf 464}, 697 (2010).

\bibitem{Teufel} J. D. Teufel, T. Donner, Dale Li, J. H. Harlow, M. S. Allman, K. Cicak, A. J. Sirois, J. D. Whittaker, K. W. Lehnert, and R. W. Simmonds, Sideband cooling of micromechanical motion to the quantum ground state, Nature {\bf 475}, 359 (2011).

\bibitem{Chan} J. Chan, T. P. Mayer Alegre, Amir H. Safavi-Naeini, Jeff T. Hill, Alex Krause, Simon Groeblacher, Markus Aspelmeyer, and Oskar Painter, Laser cooling of a nanomechanical oscillator into its quantum ground state, Nature {\bf 478}, 89 (2011).

\bibitem{Regal} R.W. Peterson, T.P. Purdy, N.S. Kampel, R.W. Andrews, P.L. Yu, K.W. Lehnert, C.A. Regal, Laser Cooling of a Micromechanical Membrane to the Quantum Backaction Limit, Phys. Rev. Lett. {\bf 116}, 063601 (2016).


\bibitem{LPNQDs} N. Somaschi,	V. Giesz,	L. De Santis,	J. C. Loredo,	M. P. Almeida,	G. Hornecker,	S. L. Portalupi,	T. Grange,	C. Ant\'on,	J. Demory,	C. G\'omez,	I. Sagnes,	N. D. Lanzillotti-Kimura,	A. Lema\^itre,	A. Auffeves,	A. G. White,	L. Lanco, and P. Senellart, Near-optimal single-photon sources in the solid state, Nature Photonics {\bf 10}, 340–345 (2016).


\bibitem{Restrepo} J. Restrepo, C. Ciuti, and I. Favero, Single-Polariton Optomechanics, Phys. Rev. Lett. {\bf 112}, 013601 (2014).

\bibitem{Kyriienko} O. Kyriienko, T. C. H. Liew, and I. A. Shelykh, Optomechanics with Cavity Polaritons: Dissipative Coupling and Unconventional Bistability, Phys. Rev. Lett. {\bf 112}, 076402 (2014).

\bibitem{Rozas_Polariton} G. Rozas, A. E. Bruchhausen, A. Fainstein, B. Jusserand, and A. Lema\^itre, Polariton path to fully resonant dispersive coupling in optomechanical resonators, Phys. Rev. B {\bf 90}, 201302(R) (2014).





\bibitem{Karrai} C. Metzger, and K. Karrai, Cavity cooling of a microlever, Nature {\bf 432}, 1002 (2004)

\bibitem{Kippenberg} T. J. Kippenberg, H. Rokhsari, T. Carmon, A. Scherer, K. J. and Vahala, Analysis of Radiation-Pressure Induced Mechanical Oscillation of an Optical Microcavity, Phys. Rev. Lett. {\bf 95}, 033901 (2005).

\bibitem{Zhao} C. Zhao, L. Ju, H. Miao, S. Gras, Y. Fan, and D. G. Blair, Three-Mode Optoacoustic Parametric Amplifier: A Tool for Macroscopic Quantum Experiments, Phys. Rev. Lett. {\bf 102}, 243902 (2009).

\bibitem{Grudinin} I. S. Grudinin, H.  Lee,  O. Painter, and K. J. Vahala, Phonon Laser Action in a Tunable Two-Level System, Phys. Rev. Lett. {\bf 104}, 083901 (2010).


\bibitem{FainsteinPRL2013} A. Fainstein, N. D. Lanzillotti-Kimura, B. Jusserand, and B. Perrin, Strong Optical-Mechanical Coupling in a Vertical GaAs/AlAs Microcavity for Subterahertz Phonons and Near-Infrared Light, Phys. Rev. Lett. {\bf 110}, 037403 (2013).





\bibitem{JMGerard_pillarDispersion} J. M. G\'erard, D. Barrier, J. Y. Marzin, R. Kuszelewicz, L. Manin, E. Costard, V. Thierry-Mieg, and T. Rivera, Quantum boxes as active probes for photonic microstructures: The pillar microcavity case, Appl. Phys. Lett. {\bf 69}, 449 (1996).


\bibitem{GayralGerard} T. Rivera, J.-P. Debray, J. M. G\'erard, B. Legrand, L. Manin-Ferlazzo, and J. L. Oudar, Optical losses in plasma-etched AlGaAs microresonators using reflection spectroscopy, Appl. Phys. Lett.  {\bf 74}, 911 (1999).



\bibitem{Thomsen} C. Thomsen, H. T. Grahn, H. J. Maris, and J. Tauc, Surface generation and detection of phonons by picosecond light pulses, Phys. Rev. B {\bf 34}, 4129 (1986).

\bibitem{Bartels} A. Bartels, T. Dekorsy, H. Kurz, and K. Koehler, Coherent Zone-Folded Longitudinal Acoustic Phonons in Semiconductor Superlattices: Excitation and Detection, Phys. Rev. Lett.{\bf 82}, 1044 (1999).





\bibitem{JusserandPRL2015} B. Jusserand, A.N. Poddubny, A.V. Poshakinskiy, A. Fainstein, and A. Lemaitre Polariton Resonances for Ultrastrong Coupling Cavity Optomechanics in GaAs/AlAs Multiple Quantum Wells, Phys. Rev. Lett. {\bf 115}, 267402 (2015).


\bibitem{JMGerard_pillardynamics} H. Thyrrestrup, E. Y\"uce, G. Ctistis, J. Claudon, W. L. Vos, and J. M. G\'erard, Differential ultrafast all-optical switching of the resonances of a micropillar cavity, Applied Physics Letters {\bf 105}, 111115 (2014).

\bibitem{RuelloUltrasonics} P. Ruello and V. E. Gusev, Physical mechanisms of coherent acoustic phonons generation by ultrafast laser action, Ultrasonics {\bf 56}, 21 (2015).



\bibitem{Baker} C. Baker, W. Hease, Dac-Trung Nguyen, A. Andronico, S. Ducci, G. Leo, and I. Favero, Photoelastic coupling in gallium arsenide optomechanical disk resonators, Optics Express {\bf 22}, 14072 (2014).

\bibitem{Rakich} P. T. Rakich, C. Reinke, R. Camacho, P. Davids, and Z. Wang, Giant Enhancement of Stimulated Brillouin Scattering in the Subwavelength Limit, Physical Review X {\bf 2}, 011008 (2012).



\bibitem{JMGerard} J. M. G\'erard, B. Sermage, B. Gayral, B. Legrand, E. Costard, and V. Thierry-Mieg, Enhanced Spontaneous Emission by Quantum Boxes in a Monolithic Optical Microcavity, Phys. Rev. Lett. {\bf 81}, 1110 (1998).

\bibitem{Rozas} G. Rozas, M. F. Pascual Winter, B. Jusserand, A. Fainstein, B.  Perrin, E. Semenova, and A. Lema\^itre, Lifetime of THz Acoustic Nanocavity Modes, Phys. Rev. Lett. {\bf 102}, 015502 (2009).


\bibitem{Painter_Transient} S. M. Meenehan, J. D. Cohen, G. S. MacCabe, F. Marsili, M. D. Shaw, and O. Painter, Pulsed Excitation Dynamics of an Optomechanical Crystal Resonator near Its Quantum Ground State of Motion, Physical Review X {\bf 5}, 041002 (2015).

\bibitem{Splitting} Note that the apparent splitting of the 19~GHz mode at $\sim 2000$~ps  (at the position of the vertical dashed line in the bottom panel of Fig.~\ref{fig02}) is a mathematical effect of the windowed Fourier transform, arising due to the $\pi$ change of phase of the mechanically induced reflectivity oscillations observed upon crossing of the cavity mode with the maximum of the laser energy. This feature can be in fact used as a signature for the dynamical transition of the cavity-mode from blue to red-shift respect to the laser maximum.


\bibitem{Marquardt2006} F. Marquardt, J. G. E. Harris, and S. M. Girvin, Dynamical Multistability Induced by Radiation Pressure in High-Finesse Micromechanical Optical Cavities, Phys. Rev. Lett. {\bf 96}, 103901 (2006).

\bibitem{Metzger} C. Metzger, M. Ludwig, C. Neuenhahn, A. Ortlieb, I. Favero, K. Karrai, and F. Marquardt, Self-Induced Oscillations in an Optomechanical System Driven by Bolometric Backaction, Phys. Rev. Lett. {\bf 101}, 133903 (2008).


\bibitem{TalCarmon} T. Carmon, H. Rokhsari, L. Yang, T. J. Kippenberg, and K. J. Vahala, Temporal Behavior of Radiation-Pressure-Induced Vibrations of an Optical Microcavity Phonon Mode, Phys. Rev. Lett. {\bf 94}, 223902 (2005).

\bibitem{Vahala2007} T. Carmon, M. C. Cross, and K. J. Vahala, Chaotic Quivering of Micron-Scaled On-Chip Resonators Excited by Centrifugal Optical Pressure, Phys. Rev. Lett. {\bf 98}, 167203 (2007).











\end{references}
\end{document}